# Light-induced giant and persistent changes in the converse magnetoelastic effects in Ni/$BaTiO_3$ multiferroic heterostructure

*Anita Bagri[1], Anupam Jana[1], Gyanendra Panchal[2], Rakhul Raj[1], Mukul Gupta[1], V.R. Reddy[1], Deodatta Moreshwar Phase[1] and Ram Janay Choudhary[1*]*

*[1]UGC-DAE Consortium for Scientific Research, Indore, India, 452001,*

*[2]Department Methods for Characterization of Transport Phenomena in Energy Materials, Helmholtz-Zentrum Berlin für Materialien und Energie, Berlin, Germany,14109*

*Corresponding Author: *ram@csr.res.in*

*Keywords: Piezoelectrics, Magnetoelastic coupling, Domain wall motion, Magnetic domains, Light-controlled magnetization,*

**Abstract:**

Magnetoelastic and magnetoelectric coupling in the artificial multiferroic heterostructures facilitate valuable features for device applications such as magnetic field sensors and electric write magnetic-read memory devices. In a ferromagnetic/ferroelectric heterostructures, the strain mediated coupling exploits piezoelectricity/electrostriction in ferroelectric phase and magnetostriction/piezomagnetism in ferromagnetic phase. Such verity of these combined effect can be manipulated by an external perturbation, such as electric field, temperature or magnetic field. Here, we demonstrate the remote-controlled tunability of these effects under the visible, coherent and polarized light. The combined surface and bulk magnetic study of domain-correlated Ni/$BaTiO_3$ heterostructure reveals that the system is strong sensitive about the light illumination via the combined effect of converse piezoelectric, magnetoelastic coupling and converse magnetostriction. Well-defined ferroelastic domain structure is fully transferred from a tetragonal ferroelectric to magnetostrictive layer via interface strain transfer during the film growth. The visible light illumination is used to manipulate the original ferromagnetic microstructure by the light-induced domain wall motion in ferroelectric, consequently the domain wall motion in the ferromagnetic layer. Our findings mimic the attractive remote-controlled ferroelectric random-access memory write and magnetic random-access memory read application scenarios, hence, can be proven as a novel perspective for room temperature device applications.

## Introduction:

Multiferroic materials exhibits at least two types of ferroic properties and the corresponding order parameters are coupled to each other, which promotes the additional functionalities that make them unique and appealing [1-2]. However, the low transition temperatures of single phase multiferroics limits them from room temperature (RT) application prospects, moreover the room-temperature multiferroicity is limited to $BiFeO_3$ and its derivative compounds [1]. An alternate strategy to engineer multiferroics is to introduce an indirect coupling between two materials such as a ferroelectric and a ferromagnetic, named as artificial multiferroics. Among these materials, RT ferroelectric lead-free $BaTiO_3$ (BTO) based artificial multiferroics are utilized for device application based on its well-known ferroelectric and piezoelectric properties [3]. In the last couple of years, much focus has been paid to the visible light induced tunability in the structural as well as the ferroic properties of BTO by the material science experimental as well as theoretical community [4-7]. The significant discovery of photostriction as well as photopolarization in $BaTiO_3$ has been recognized as a revolution in remote controlled based technology, which has made the task of remote controlling of ferroic properties an appealing topic in Materials Science [8-11]. The ferroelectric materials exhibit coupled degrees of freedom, such as natural occurrence of switchable polarization coupled to the strain. The ability of transformation of the optical energy to the mechanical strain of this classic piezoelectric may offer the remote-controlled functionality of optomechanical devices e.g., light controlled elastic micromotors, micro actuators, sensors, photoacoustic devices, etc as like the earlier discovered piezoelectrics [8,12-14].

Besides the light-controlled ferroelectricity, this discovery certainly attracts the light controlled multiferroicity in the BTO-based artificial multiferroic and magnetoelectric heterostructures. From the application point of view, each layer of ferroelectric-ferromagnetic heterostructure allows a separate promising tuning of the relevant physical properties, such as structural, electrical or magnetic properties, manifesting the potentiality of these novel devices. These multilayer structures have been proven as an add-on to the low-power spintronic devices and magnetic random-access memories (MRAM) [15-17]. In such heterostructures, the structural, electrical or magnetic properties are intertwined and can be tuned either via electric biasing across the ferroelectric substrate or magnetic field. The initial studies of these artificial multiferroic heterostructures under the electrical biasing reveals the feasibility to write the giant, sharp and persistent magnetic changes, attracts the low power MRAM device applications [18]. To date, the tuning of magnetic properties of any ferromagnetic-ferroelectric hybrid structure is reported with the external parameter such as electric field control [19-27], temperature control [20, 24-26], and magnetic field control [20,23,25,27]. This external perturbation comprises a heavy electrical circuitry and appeals for a weight reduction manipulator. The remote control of magnetization without varying temperature or magnetic field or electric field represents a higher scientific goal. The remote controlling of the magnetic properties of such hybrid heterostructures is possible with the converse magnetostrictive route. Some of the $BiFeO_3$-based heterostructures are reported, where the magnetic properties of the top-grown magnetic layer are tunned via light-induced changes in the lattice constant of the $BiFeO_3$ substrate [28-30]. In contrast, evaluating and understanding such kind of tunability in the microscopic magnetoelectric and magnetoelastic coupling via the light-induced modification in BTO-based ME heterostructures has not been reported so far. Such remote controlled tunability in the magnetism of this material combination can be proven as a novel perspective for room temperature device applications.

In this work, we have demonstrated that it is possible to precisely write the regular ferromagnetic domain patterns and control the motion of magnetic domain walls in BTO-based artificial heterostructures via illumination of light. Our approach consists of three steps. First, well-defined ferroelastic domain structure is fully transferred from a tetragonal ferroelectric to magnetostrictive layer via interface strain transfer during the film growth. Hereafter, the magnetoelastic coupling at the interface of the ferroelectric-ferromagnetic material leads to change in the magnetic anisotropy in the different structural phases of the BTO. In the third step, visible light illumination is used to manipulate the ferromagnetic microstructure by the light-induced domain wall motion in ferroelectric. To demonstrate this, the 40-nm-thick magneto-strictive nickel (Ni) film deposited onto the single crystal barium titanate (BTO) substrate. In the Ni film, the substrate induced strain is tuned by illuminating the BTO substrate by visible light of 532 nm wavelength, hence varying the substrate lattice constant as well as the ferroelectric domains contribution owing to the photostriction and photo-domain properties of $BaTiO_3$. The structural change as well as the domain switching in the BTO single crystal is confirmed by the x-ray diffraction (XRD) measurements under the light illumination. It is observed from magneto-optic Kerr microscopy (MOKE) study that the ferroelastic domains of BTO are fully imprint on the magnetic domains of the Ni as well as the ferroic domains boundaries are pinned. Light triggers the domain wall motion in the BTO, responsible for the magnetic domain wall motion in the ferromagnetic layer, which significantly changes the magnetization behaviour of Ni layer. Consequently, under the light illumination, a giant and persistent magnetoelastic effects are observed in Ni/$BaTiO_3$ heterostructure. Our findings in the context of such remote controlled tunability in the magnetism of this material combination can be proven as a novel prospective for spintronics device applications.

## Results and discussion:

### *(a) Structural and Topographic characterization:*

The θ/2θ x-ray diffraction (XRD) patterns of Ni/BTO, Ni/LSAT, Ni/Si films are shown in **Fig.1 (a),** From the XRD patterns of the grown films on (001) LSAT and (001) Si substrates depicts the (111) orientation of grown films of Ni, confirming the crystalline nature of grown film. On the Si substrate besides (111) orientation of Ni, (002) reflection is also present with less than 10% intensity of the preferred orientation. In our study, we are not able to detect the preferred orientation of Ni-film on BTO substrate, whereas the simultaneously Ni-film grown on (001) LSAT film shows the (111) preferred orientation (see **Fig. 1(b)**). XRD pattern of BTO single crystal reveals Bragg reflections corresponding to (*00l*) and (*h00)/(0k0*) owing to its tetragonal symmetry at room temperature. We designate (*00l*) peak arising due to *c*-domains and (*h00)/(0k0*) to the *a*-domains. The preferred orientation of Ni (111) peak position matches exactly with the shoulder of the (002) reflection of out-of-plane polarized *c*-domains and hence we do not observe a distinct reflection from the Ni film on the BTO substrate. In the chosen BTO single crystal substrate, the higher relative intensity of the *a*-domain with the reflection (*h00/0k0*) suggests the dominance of the *a*-domain in crystal.

To confirm the growth of Ni layer on BTO crystal, we have done the atomic force microscopic measurement to check the morphology of our Ni/BTO hybrid structure in a constant height mode. The observed morphology of the Ni layer in $2 \times 2$ $\mu m^2$ window is shown in **Fig. 1(c)**. A careful look of the island type growth of Ni layer suggests that the average heights (*h*) of constituting grains are ~ 40 nm with the average lateral length (*l*) of 50 nm respectively. The cumulative investigation of grown films of Ni on LSAT and Ni as well as the lateral growth information of Ni on BTO substrate confirms the crystalline nature, perhaps the very slow growth rate as well as the coexistence of the (111) and (002) orientation of Ni and BTO respectively at same $2\theta$ value, mislead the structural information of Ni/$BaTiO_3$ heterostructure.

## *Ferromagnetic-ferroelastic correlations:*

Full imprinting of ferroelastic $BaTiO_3$ domains into Ni during the film growth is demonstrated in **Fig. 2 (a)**. In Fig. 2(a), the microscopic image of the Ni/BTO film at 0 Oe magnetic field is mapped using the polarized light, in which the dark and white contrasts appear due to distribution of ferroelectric BTO domains, mostly a and c domains, having different polarization directions. The images recorded at ± 20 Oe magnetic field (which is less that saturating magnetic field of the Ni layer) are its corresponding MOKE images revealing the magnetic contrast of the grown ferromagnetic Ni layer on the BTO substrate. Analysis of birefringent contrast of ferroelectric $BaTiO_3$ substrate and magneto-optic Kerr contrast of ferromagnetic Ni layer indicates that the ferroelectric and magnetization directions are collinear in the absence of applied field. The precise transfer of the ferroelastic domain structure of the underlying BTO substrate as the magnetic domains in the Ni film can be explained in terms of the strain transfer at the Ni/BTO heterostructure interface, which induces the modulations in the magnetoelastic anisotropy via inverse magnetostriction.

Comparing the ferroelastic and ferromagnetic domains of BTO and Ni layer respectively, interestingly, it is observed that the ferromagnetic domains of Ni layer is strong magnetoelastically coupled with the ferroelastic domains of BTO and shows that the magnetization of Ni follows up the imprint of the polarization of BTO crystal. Resultantly, the Ni layer grown onto the selected *a* and *c*- dominated domains region of BTO shows soft and hard MOKE loop (**see Fig. 2(b)**) with the coercivity ($H_c$) of 17 and 34 Oe respectively.  In the *a*-domains, the magnetic moments switch abruptly when the field is reversed, while in the *c*-domains, the magnetic moments rotate coherently towards the easy axis when the magnetic field is applied. Besides this, it is observed that the ferromagnetic domain wall is pinned at the ferroelastic domain boundaries, observed upon the application of magnetic field.

The MOKE loop of Ni/BTO structure, which is shown in the right part of the **Fig. 2 (b),** clearly manifests the superposition of the MOKE loops, recorded at the domains, which have the dominance of *a* and *c* domains, consisting the dual hysteresis. The combination of strong domain pinning at ferroelectric domains' boundaries and distinctive magnetization reversal process in the neighbouring strip domains provide a unique functionality such as domain wall magnetoresistance.

## *Magnetoelastic coupling in Ni/BaTiO$_3$ heterostructure:*

To realize the mechanical strain induced modifications in the magnetic properties of the Ni film, we carried out Magnetization versus temperature (M-T) behaviour of Ni thin film deposited on BTO substrate in zero field cooled (ZFC) and field cooled warming (FCW) protocols in the temperature range of 5 to 350 K at 50 Oe magnetic fields as shown in **Fig. 3(a)**. We have applied the magnetic field in the (100) direction of the Ni/BTO sample, which is along the length of the magnetic domains as shown in the schematic of **Fig.3**. We observe a discontinuous jump by 82 % in the magnetization of the Ni film at temperature 194 K, where the rhombohedral (R) to orthorhombic (O) (R $\rightarrow$ O) structural transition of BTO takes place. Interestingly, when the temperature reaches to the orthorhombic (O) to tetragonal (T) (O $\rightarrow$ T) phase transition temperature at 285 K, the magnetization drops by 15 %. The value of % (increase) or drop (decrease) is defined by $[(M_f - M_i) / M_i] \times 100$, where $M_i$ and $M_f$ are the magnetization values just before and after the transition in the FCW cycle. It should be noted that such abrupt changes in the magnetization are not the characteristics of the Ni thin film. One can see the M-T behaviour of Ni thin film grown on (001) LSAT substrate under a similar condition does not show the sharp discontinuity as shown in the inset of **Fig. 3(a)**. M-H loop study in these individual BTO phase regime manifests a huge change in the coercivity as well as in the loop shape of the Ni/BTO heterostructure (see **Fig. 3(b)**). This clearly highlights that the magnetic properties of Ni film are strongly coupled with the BTO structural transition and reveals the mechanical strain induced modifications in the magnetic properties of the Ni film via the magnetoelastic coupling between the Ni and BTO.

In a ferromagnetic/ferroelectric heterostructures, the strain mediated coupling exploits piezoelectricity/electrostriction in ferroelectric phase and magnetostriction/piezomagnetism in ferromagnetic phase. Since ferroelectric BTO forms a multidomain state upon crossing the phase transitions, the non-uniform strain state is induced into the overlying ferromagnetic layer, consequently the magnetic properties are changed. Earlier, Gerprags *et al* [20], reported that the % jump or dip in the magnetization of Ni/BTO hybrid structure across the BTO structural phase transition temperature strongly depends upon the volume fractions of different ferroelastic domains during deposition or the initial state of the measurements. In our study, the observed percentage of magnetization jump to 82 % at the orthorhombic to rhombohedral phase transition temperature is larger than earlier reports [20], which discard the possibilities of any mechanical fatigue due to BTO. Overall, the magnetic study of Ni/BTO heterostructure in the absence of light illumination reveals that the magnetic properties of Ni are strongly magneto elastically coupled with the BTO.

The substrate induced strain through the structural transition of BTO and temperature dependent structural change of Ni layer can be discussed via the substrate clamping effect. From the temperature dependent lattice constant behaviour of BTO, on cooling from T-phase to R-phase, $a_T^{300\,K} \times a_T^{300\,K}$ template (*c*-domains) shows the in-plane isotropic expansion, whereas the $a_T^{300\,K} \times c_T^{300\,K}$ template (*a*-domains) produces a nominally uniaxial contraction of ~1%. The magnetoelastic energy for any magnetostrictive material is defined by [31], $K_{ME} = -\frac{3}{2}\lambda\sigma Cos^2\theta$ , For Ni, magnetostriction constant $\lambda$ is negative and therefore, $\sigma > 0$ (tensile stress) favours $\theta = \pi/2$ *i.e.*, moment alignment is perpendicular to the stress axis, whereas $\sigma < 0$,(compressive stress) fevours $\theta = 0$ *i.e.*, moment alignment in the stress direction. Therefore, the negative magnetostrictive Ni layer rotates its easy axis in the in-plane direction having the maximum contraction. Accordingly, the temperature dependent magnetization behaviour shows the drop and jump in M-T behaviour of Ni/BTO structure across the T to O and O to R structural transitions of the BTO substrate respectively, when the magnetic field is applied in (100) direction of BTO crystal.

To further explore this switching or reorientation of magnetization phenomena, we have performed the angle dependent magnetization versus applied magnetic field (M-H) measurements along the three axes of the Ni/BTO sample as shown in the schematic of **Fig.3** at three different temperature values of 100, 230, and 300 K corresponding to rhombohedral (R), orthorhombic (O), and tetragonal (T) phases of BTO respectively as shown in **Fig. 3(c)**. With decreasing temperature, the magnetic anisotropy of Ni/BTO heterostructure changes and huge enhancement in coercivity is observed. When the system enters from T to O phase, the coercivity enhances from 40 Oe at 300 K (T-phase) to 160 Oe at 230 K (O-phase), subsequently 380 Oe at 100 K (R-phase). From the angle dependent M-H loop study clearly reveals the strain-induced lattice distortion due to successive phase transition of BTO modifies the magnetic anisotropy and easy axis rotates owing to magnetoelastic coupling. Such strong magnetoelastic coupling has been reported for other hybrid structure such as, $La_{1-x}Sr_xMnO_3$/BTO, where the magnetic anisotropy follows fourfold to twofold to almost isotropic behaviour, when the BTO passes through T to O to R phase transitions, respectively [25].

When the Ni/BTO system undergoes from T→O→R phase of BTO, it is noted that from the angle dependent M-H loop, among the three sample axes, (010) axis is the magnetic easy axis of Ni/BTO heterostructure in the R-phase regime of the BTO, which switches to (100) as an easy axis in the O-phase regime of the BTO. In the T-phase regime of BTO, the magnetic behaviour of Ni/BTO becomes almost isotropic in the plane of sample, whereas strong anisotropy is revealed in the (001) direction. As the system undergoes from T to O phase of BTO, the easy axis switches to (100) direction, which becomes hard in the R-phase, resulting jump and drop in the magnetization in O and R-phase regime of BTO respectively, when the magnetic field is applied along the (100) direction, as observed in the M-T behaviour. Overall, the magnetic study of Ni/BTO heterostructure suggests that the magnetic properties of Ni are strongly magneto elastically coupled with the BTO structural transition. Therefore, such strong magnetoelastically coupled Ni/BTO can be a good candidate to demonstrate the magnetization tunability under the light illumination.

## ***Photostriction and Photodomain effect in Ni/BaTiO$_3$ heterostructure:***

In order to investigate structural changes under visible light illumination, we have recorded the XRD patterns while shining the monochromatic, coherent, and polarized laser light of wavelength of $\lambda\sim$ 532 nm and power of 10 mW onto to the Ni/BTO heterostructure as shown in the **Figure 4**. For a detailed investigation, we have selected the most intense Bragg's reflection lines of BTO, which are (002) and (200/020), corresponding to *c* and *a*-domains respectively. The splitting of each reflection is corresponding to the Cu $K_{\alpha1}$ ($\lambda$=1.5406 Å) and $K_{\alpha2}$ ($\lambda$=1.5444 Å) radiation of the x-ray source. The step size during the XRD scan was kept at 0.002°, with the acquisition time 0.1 sec It is clearly observed that under the light illumination, the (002) line is shifted significantly towards the higher 2θ value [from 44.879° to 44.896° ($K_{\alpha 1}$ line)], whereas the peak (200/020) does not show any observable deviation [~ 45.402°($K_{\alpha 1}$ line)]. The estimated out-of-plane lattice parameter (L.P.) of *c*-domain (c) is found to be 4.036 ± 0.001 Å and 4.034 ± 0.001 Å in the absence and presence of the illumination respectively, which amounts to a 0.05 % compressive photo strain. The out-of-plane L.P. of *a*-domain (a) remains unchanged with light and is found to be 3.991 ± 0.001 Å. Overall, the tetragonality of BTO is decreased from 1.011 to 1.010 with the illumination of light. Besides the photostrain, the 14% *a* to *c*-domain switching is observed for the present BTO single crystal (calculated from the XRD integrated intensity ratio of a and c domains), which is consistent with the earlier reports [5]. The modulation in the lattice constants of BTO allows us to induce a strain in the nickel film by illuminating the BTO substrate by visible light during the measurements. We shall like to mention here that these observations are reproduced in the bare single crystal of BTO (**supplementary section S1)**.

Moreover, under the light illumination, there is no peak broadening in the out-of-plane lattice parameters of the *a* and *c*-domains of the BTO substrate, which discards the possibilities of light-induced nanoscale inhomogeneities. Besides this, no significant sample heating is noted from the x-ray scan during the laser illumination as this would yield the lattice expansion, contrary to our findings. The observed contraction along the (002) direction will lead to the overall lattice expansion in the in-plane lattice parameters of the *c*-domains**.** The light is therefore able to induce anisotropic deformation in BTO, that can be used to transfer stress into the magneto-strictive overlayer.

## *Influence of photostriction and photodomain effect onto the magnetic properties of Ni/BTO heterostructure:*

The strain mediated change of magnetic properties of Ni layer with the ferroelectric BTO crystal suggests that the light illumination also might influence the magnetic properties of Ni layer via the combined effect of converse piezoelectric, magnetoelastic coupling and converse magnetostriction.

Thus, we have carried out the magnetic measurement of Ni/BTO heterostructure with and without the visible light illumination in the 7 T SQUID-VSM to investigate the effect of light-induced modulation in the magnetization of Ni film.

The *in situ* visible light illumination was incident parallel to the Ni/BTO heterostructure along to the length of the magnetic domains *i.e.* (100) direction of the sample, as shown in the **Schematic I**. M-T behaviour in ZFC (not shown here) and FCW protocols under the 50 Oe applied magnetic field along the in-plane (100) direction with and without visible light illumination are shown in **Fig. 5(a)** with continuously monitoring the temperature fluctuations.

It is evident that upon the light illumination, the magnetization is drastically enhanced by ~15% in the R-phase regime, it is reduced considerably in the O-phase regime (by ~10%) and remains invariant in the T-phase regime of BTO. In order to check the influence of light on the structural transition of BTO, we have taken the first order derivative of each FCW MT curve, which are shown in the **Fig. 5(b)** and **(c).** In the first order derivative, the peak and dip are observed corresponding to the sudden jump and drop in the magnetization versus temperature curve indicating the R$\rightarrow$O and O$\rightarrow$T structural phase transition of BTO via the sudden change in the magnetization of Ni layer. Under the light illumination, it is observed from the dM/dT behaviour of Ni/BTO structure, R$\rightarrow$O transition temperature shifts from 194.50 to 196.50 K, whereas the O$\rightarrow$T transition temperature shifts from 285 to 286.3 K. This clearly manifests the strong influence of light on the magnetoelastic coupling of Ni/BTO hybrid structure. The enhancement in the structural transition temperatures in the present study is in contrast to the recent study by Dwij *et al.* [10] where they have shown the shift in the structural phase transition of pure BTO crystal towards lower temperature due to the presence of photo-strain effect through the temperature dependent dielectric study of BTO crystal.

## *Giant and persistent piezomagnetism in Ni/BTO heterostructure via the light-induced changes in the magnetoelastic coupling*

In order to further probe the influence of light on the magnetoelastic coupling of Ni/BTO structure, we have investigated the light switching effect on the light induced changes on the magnetization of Ni/BTO heterostructure recorded at different magnetic field values of 100 and 400 Oe at 100 K, as shown in **Fig. 6(a) and (b)** respectively. Before recording any measurement, we have demagnetized the system via heating process up to 360 K, which is beyond the Curie temperature of Ni film, to discard the possibility of existence of remanence before any recorded measurement. Here we want to mention that 400 Oe value is chosen such that the applied magnetic field value is above the coercivity of Ni layer at 100 K and lower than the magnetic field at which the system gets saturated.

All the measurement of magnetization evaluation with time are done along the (100) axis, which is not the easy axis for Ni layer in rhombohedral phase of BTO at 100 K. In the rhombohedral phase regime of BTO, it is observed that with the illumination of light, the magnetization of Ni layer enhances with time, reproducibly observed at different field 100, and 400 Oe. It is observed that during the dark state, the magnetization is saturated, i.e., remains fairly constant for the particular applied magnetic field, and during the On condition, the magnetization enhances successively. For magnetic field values of 100 and 400 Oe, we have recorded the magnetization data under the illumination five cycles of light pulse. We have estimated the 10 and 2 % of enhancement in the magnetization of Ni/BTO heterostructure at 100 K, at 100 and 400 Oe applied magnetic field respectively via the photo-illumination of heterostructure. The decrease in the percentage of enhanced magnetization at 400 Oe as compared to that of 100 Oe is expected, since the coercivity at 100 K is 300 Oe and hence the domain wall motions in the Ni layer is not so prominent at 400 Oe as compared to that at 100 Oe. It is also observed that the rate of enhancement in magnetization of Ni layer upon successive light illumination decreases non linearly for each field as shown in the **Fig. 6(c)**. Such signature of magnetization enhancement of Ni/BTO heterostructure at room temperature is also observed as the modified dichroic signal under the light illumination in the x-ray circular magnetism dichroism study shown in the **Fig. 6(d)**. The magnetization enhancement in the ON condition clearly discards the thermal effect, as in such cases the local rise in temperature will cause a decrement in the magnetization. In all the photo-magnetic cycles, the effect of light on the domain wall of Ni layer does not appear to be a transient phenomenon, rather it appears as a sustainable state, i.e., the magnetic state is irreversibly modified upon light illumination.

## *Discussion*

The polarization microscopy analysis of both the ferroic domains structures reveals that the ferroelastic and ferromagnetic microstructures are strongly coupled to each other via converse magnetoelastic effects. Using the hard axis magnetization MOKE loop, we have estimated the magnitude of the magnetoelastic anisotropy of alternating $a$ and $c$ domains as ~ 1.60 $\times 10^4$ J/m$^3$. This anisotropy can be compared with the theoretical estimation. For the polycrystalline magnetostrictive materials, the magnetoelastic energy is given by the expression [31]

$$K_{ME} = -\frac{3}{2}\lambda\sigma Cos^2\theta$$

Where, $\lambda$ is the magnetostriction constant, $\theta$ is the angle between the spontaneous magnetization and stress directions, $\sigma$ is the stress, which is proportional to the strain $\varepsilon$ via Young's modulus Y. Considering the full strain transfer from $BaTiO_3$, which is estimated from the room temperature mismatch between in-plane ($a$) and out-of-plane ($c$) lattice constant of $BaTiO_3$ [21], *i.e.*, $\varepsilon$ (%)= $\{(c - a)/a\} \times 100$ = 1.1% to the overlying Ni layer and λ = -3.55 × $10^{-5}$ [32], Y= 2.05× $10^{11}$ J/m$^3$ [33] for Ni layer, gives $K_{ME,Max}$ = 1.10 × $10^5$ J/m$^3$. This estimation suggests that BTO does not transfer the full strain to the Ni film, rather it is less than 10% of the full lattice strain. Yet, the MOKE experiments clearly evidence that such small strain effect is sufficient to fully imprint the ferroeleastic domains into magnetostrictive films. Another strong observation in the MOKE images of Ni/BTO is the magnetic domain wall pinning to the domain wall boundaries of underlying ferroelectric BTO substrate with the potential barrier of 20 Oe magnetic field. Such pinning plays a crucial role in the cooperatively motion of ferroelectric and ferromagnetic domain wall under the light illumination.

Very recently, Rubio-Macros *et al* [4-5,9] reported the domain wall motion in $BaTiO_3$ single crystal as well as polycrystalline samples upon light irradiation. The unexpected coupling between the light and ferroelectric polarization modifies the stress at the domain wall, leading to the rearrangement of domains. The accumulated charges at the domain wall create an asymmetric saw-teeth potential[34-36], giving rise the so-called ratchet effect[37-38], which causes the domain wall motion. Resultantaly, photostrain, domain switching and macroscopic changes in polarization are observed. In the present case of Ni/BTO, as we have demonstrated fully imprint of the ferroelastic domains into the ferromagnetic domains of the Ni film. It is also shown above that the magnetization is modified upon laser illumination owing to the photo-strain effect of the underlying BTO substrate, pointing out towards the converse magneto-striction effect in the film. In such artificial ferroic-coupled system, under the light-illumination, the ferroelectric domain wall motion (write) can be recorded as the magnetic domain wall motion (read), hence the magnetization of ferromagnetic layer changes. The domain wall motion as well as the domain reorientation in the BTO crystal using the polarized visible light is well described in the literature. From the MOKE image analysis of ferromagnetic domain patterns of Ni/BTO, it is revealed that under the light illumination of 405 nm with the power of 10 mW, not only the ferroelastic domain wall moves, ferromagnetic domain wall also moves (see the **Fig. 7**). Ferroelastic domain switching under the light illumination redistributes the fraction of soft and hard stripy ferromagnetic domains in the Ni layer, consequently the magnetic properties of Ni/BTO are modified as observed from the magnetization versus temperature behaviour under the light illumination.

It is important to note that the light induced ferroelectric domain wall motion can lead to two possibilities (i) the change in the volume fraction of the ferroelastic domains, and hence (ii) change in the strain state in each structural phase. From the room temperature XRD patterns, the 0.05 % photostrain on $c$-domains as well as the 14% $a$ to $c$ domains switching under the visible light illumination are evidenced. The earlier experimental observations [24] of temperature dependent magnetic imaging of Ni/BTO hybrid structure using the photoemission electron spectroscopy (PEEM) as well as magnetic force microscopy (MFM) revealed that it is the $c$-domains which grow and $a$-domains annihilate on cooling the hybrid structure as discussed above.

Under the light illumination, at a fixed temperature below the Curie temperature, the change in magnetic energy density of the ferromagnetic layer can be written as….

$$u_{T,Light}^{FM} - u_0 = \Delta u_{ani}^{FM}\,(\boldsymbol{m}_{Ni}) + \Delta u_{el}^{FM}\left(\eta_k^{Ni}\right) + \Delta u_{ME}^{FM}\left(\boldsymbol{m}_{Ni}, \eta_k^{Ni}\right) + \cdots\ldots \quad (1)$$

The first term in the Eqn. (1) denotes the change magnetic anisotropy, the second term $\Delta u_{el,}^{FM}\left(\eta_k^{Ni}\right)$ describes the change in the purely elastic energy density of Ni and is the function of the strain components. $\Delta u_{ME}^{FM}\left(\boldsymbol{m}_{Ni}, \eta_k^{Ni}\right)$ describes the change in magnetoelastic energy density and hugely depends upon the strain components ($\eta_k^{Ni}$) and the direction of magnetization $\boldsymbol{m}_{Ni}$ and represents the interaction between the elastic and magnetic anisotropy energies. The external magnetic field applies a torque to the magnetization to rotate it in the field direction, whereas the anisotropy field tends to rotate the magnetization towards the easy axis. The illumination of light is not directly influencing the magnetization of the sample, rather it leads to perturb the magnetoelastic energy via photostriction, domain wall motion, as well as domain switching effects in the underlying BTO substrate, consequently affecting the competitive torques in the ferromagnetic layer Ni. As discussed earlier, the applied field direction (100) is the hard axis in the R-phase and easy axis in O-phase of BTO of magnetization. In the M-T behaviour, under the light illumination, with the applied magnetic field (50 Oe) in the R-phase of BTO favours the ferromagnetic domain wall motion such that the fraction of the magnetic domains is aligned in the easy axis direction *i.e.,* (010) grows. However, in the O-phase of BTO, where the easy axis of magnetization is already parallel to the applied magnetic field, the reason of the suppression in the magnetization is the change in the strain due to the enhanced fraction of $a_T^{300\,K} \times a_T^{300\,K}$ templates, which favours the magnetization in the out-of-plane direction. The shift in the R-O and O-T structural transition temperature towards the higher temperature gives clear evidence of change in the strain state in the different phases of BTO via photostriction and photodomain effect.

The evolution of magnetization of Ni/BTO structure in R-phase (at 100 K) under the light illumination is further confirmed from the time-evolution behaviour of magnetization at an applied fixed value of magnetization *i.e.*, 100, and 400 Oe in (100) direction, performed under the ZFC protocol. After applying a fixed value of magnetic field, the system is left to saturate its domain wall motion. After that, upon switching the laser light ON, it is observed that magnetization of system gradually increases, suggesting that the light energy triggers the domain wall motion again. After switching off the light, it is observed that the magnetic state of the system after the light illumination does not acquire the same magnetic state before the light illumination, suggesting the irreversible domain switching under the light illumination. This memory effect can be erased by the heating the sample to its paraelectric state.

We demonstrated the observation at 100 < $H_c$ (at 100 K) and 400 Oe > $H_c$ (at 100 K), following up the ZFC protocol, by switching the light in On and Off conditions. We have compared the rate of magnetization evolution ($dM/dt$) in each time interval cycle at applied magnetic field 100 and 400 Oe. It is revealed that $dM/dt$ is larger for the 400 Oe applied magnetic field at each cycle as compared to the 100 Oe applied magnetic field and non-linear for both the field values. This suggests that the domain wall motion under the light illumination saturates faster at 100 Oe as compared to the 400 Oe. Overall, the light-induced perturbation in the strain state of BTO in the different phases becomes the cause of the change in the magnetization of Ni layer.

## Conclusion:

In this paper we have clearly demonstrated the giant and persistent piezomagnetism in a ferromagnetic layer (Ni) by the means of light induced piezoelectric properties of ferroelectric $BaTiO_3$ substrate. Magnetization enhancement in this multiferroic heterostructure is driven by efficient interfacial machenical strain coupling between the Ni layer and the ferroelectric domains of $BaTiO_3$. To reorient the magnetization, a modest flux of visible light is required and the magnetoelastic coupling provides a light-controlled tunability of the ferromagnetic film. The demonstrated ability to deterministically switch the orientation of magnetization with light controlled tunability is potentially interesting for device fabrications such as light sensor, actuator and other devices that could functionally utilize light-controlled manipulation of magnetic anisotropy. Consequently, the magnetic response might also prove useful in the prospect of non-destructive, remote controlled spintronics technology.

## Methodology:

### *Sample preparation:*

To illustrate this proof-of-concept, we have chosen a high-quality commercial (100) ferroelectric $BaTiO_3$ (BTO) single crystal (3 mm × 2.5 mm × 0.5 mm) in its tetragonal symmetry (*P*4*mm*) to grow the ferromagnetic layer. In the tetragonal phase, the polar axis is aligned along one of the pseudo cubic directions, since there are six equivalent pseudo cubic, therefore six ferroelectric and three ferroelastic domains are possible. Present BTO crystal exhibits polydomain state constituted by out-of-plane polarized *c*-domain (*00l*) and in-plane polarized *a*-domain (*h00*/*0k0*) with the dominance of *a*-domains. In our study, we have chosen Ni as a magnetic layer owing to its large magnetostriction constant $\lambda \approx -3.55 \times 10^{-5}$ [31] and large magnetoelastic constants at room temperature [39], aiming the significant change in the magnetization via the light-induced perturbation in the BTO single crystal. The Ni film (40 nm) film was deposited at room temperature via dc magnetron sputtering onto the flat side of BTO single crystal and also on (001) LSAT and Si substrates simultaneously for the reference with the deposition rate of 2 Å/ sec. To protect the Ni metallic film from the oxidation we have grown a Carbon layer of 2 nm thickness.

***Structural characterizations***

X-ray diffraction (XRD) measurements were carried out using a Bruker D2 PHASER setup with Cu $K_\alpha$ lab source ($\lambda$=1.5406 Å), collecting data in θ/2θ geometry in the 10 to 90°, 2θ range for the structural characterizations of grown Ni thin films on different substrates. The step size during the XRD scan was kept at 0.002°, with the acquisition time 0.1 sec. In the present study of Ni/$BaTiO_3$ heterostructure, we considered all in-plane polarized domains as *a*-domain (*l00/0l0*) and all out-of-plane polarized domains as *c*-domain (*00l*). For a detailed investigation, we have selected the most intense Bragg's reflection lines, which are (002) and (200/020), corresponding to *c*- and *a*-domains respectively. The splitting of each reflection is corresponding to the Cu $K_{\alpha1}$ ($\lambda$=1.5406 Å) and $K_{\alpha2}$ ($\lambda$=1.5444 Å) radiation of the x-ray source. To do the experiment in the illumination condition, we have used the polarized coherent visible laser diode with the wavelength of 532 nm and of the power of 10 mW for illumination, whereas the data collected during the off condition of light illumination is considered as the dark condition. The light spot was adjusted at 0.5 mm diameter using the iris.

***Magnetic Measurements:***

7T-SQUID-Vibrating sample magnetometer (SVSM; Quantum Design, Inc. USA) was used to record the magnetization data of Ni/ $BaTiO_3$ heterostructure. The thermal demagnetization data is recorded with applying the magnetic field parallel to the pseudo-cubic [100] direction. The *in-situ* light illumination on Ni/$BaTiO_3$ heterostructure is performed in the home-made probe as shown in the **Schematic I**. We have used the monochromatic, coherent, and polarized laser light with the wavelength of $\lambda\sim$ 532 nm and power of 10 mW using the optical fiber of 0.5 mm diameter. Before using the probe for light-induced magnetic study, it is calibrated with the Teflon ball, which shows the diamagnetic character [**supplementary section S2**] The light is incident on the cross-sectional area of the Ni/$BaTiO_3$ heterostructure. The temporal evolution of magnetization is performed in the zero-field cooled protocol at 100 K (R-phase), hereafter the fixed values of magnetic field i.e., 50 Oe, 100 Oe, 400 Oe were applied along the magnetometer axis. The *in-situ* switching of light is controlled by *ex-situ* voltage supply.

***Magneto-Optic Kerr microscopy:*** To independently image the ferromagnetic and ferroelectric patterns, we have used a M/s Evico magnetic, Germany polarization microscope with an adjustable diaphragm, 100 × objective, a digital camera, equipped with a white light LED source. The measurements are performed at room temperature in longitudinal geometries for measuring the in-plane components of magnetization with the applying the external magnetic field of 0.01 Oe step size.

***Atomic force microscopic measurements:*** An atomic force microscope (AFM) was used to collect the topographic information of the surface of grown Ni layer onto the BTO single crystal. AFM measurements were performed by the Bruker Biscope Resolve set up with Silicon Nitride tip ($Si_3N_4$), in the contact mode configuration.

## Acknowledgments

We thank Dr. R.Venkatesh and M. K. Gangrade for AFM measurements. A.B. is also thankful to Rishabh Shukla for fruitful discussion.

## Conflict of Interest

The authors declare no conflict of interest.

**Figure Captions:**

**Figure 1.** ***Structural and topographic characterization of Ni thin films on different substrate*:** (a) Out-of-plane x-ray diffraction (XRD) of Ni thin films on (001) oriented $BaTiO_3$ (BTO), LSAT and Si single crystalline substrate. The different orientation crossponding to grown films and substrates is labelled with (*f*) and (*s*) respectively in the subscript of different orientations, (b) Zoomed view of θ/2θ XRD of Ni films on the BTO and LSAT substrate in the 2θ range of 42 to 52 º, (c) Topographic image of Ni film grown on the BTO substrate in a window of $2 \times 2$ $\mu m^2$ area.

**Figure 2.** ***Ferroelastic and Ferromagnetic correlations of Ni/$BaTiO_3$ heterostructure:*** (a) Polarized light and magneto-optic Kerr microscopic images of Ni/BTO heterostructure in the absence and presence of ± 20 Oe applied magnetic field respectively, correspondingly reflecting the ferroelastic and ferromagnetic domains of BTO and Ni/BTO heterostructure. (b) *a* and *c*-domain selective and averaged MOKE loops of Ni/BTO heterostructure. Red dotted line in the loops indicate the field, at which the MOKE images are shown in (a).

**Figure 3.** ***Transparent magnetoelastic coupling of Ni/$BaTiO_3$ heterostructure*:** Schematic is representing the direction schematic of Ni/BTO sample, in which the magnetization measurements are performed and the direction parallel to the length of the magnetic domains is considered as the (100) direction**,** (a) Temperature dependant magnetization behaviour (M-T) of Ni/BTO heterostructure in the zero-field (black) and field cooled (Red) protocols, when the 50 Oe magnetic field applied is applied along the (100) direction of sample. Different colour regime yellow, purple, cyan are indicating the temperature regime, in which the BTO substrate is in its rhombohedral, orthorhombic and tetragonal phase respectively. Inset shows the temperature dependent magnetization behaviour of Ni film grown on LSAT substrate. (b) M-H loops of Ni/BTO at different temperature corresponding to R

(Blue), O (Black), and T-phase (Red) of BTO substrate and the magnetic field is applied in (100) direction of the sample. (c) Angle-dependant M-H loops of Ni/BTO heterostructure in the R (at 100 K), O (at 230 K), and T-phase (at 300 K) regime of the BTO.

**Figure 4.** ***Categorical evidence of photostriction in Ni/BaTiO$_3$ heterostructure*****:** Zoomed view of θ/2θ XRD of Ni/BTO heterostructure in the 2θ range of 44.5 to 55.8 ° in the absence and presence of the 532 nm visible light illumination. The peaks crossponding to (002) and (200/020) reflections of BTO are splitted owing to the Cu $K_{\alpha 1}$ (1.5406 Å) and $K_{\alpha 2}$ (1.5444 Å) radiation of x-ray source.

**Schematic I:** Schematic diagram of the light illumination during the magnetization measurements in Vibration sample magnetometry (VSM).

**Figure 5.** ***Univocal evidence of light-induced tunability in the magnetization of magnetoelastically coupled Ni/BaTiO$_3$ heterostructure*****:** (a) M-T behaviour of Ni/BTO heterostructure in the absence and presence of 532 nm light illumination. (b) and (c) are showing the dM/dT behaviour of Ni/BTO at T→O and O→ R phase structural phase transition regime of BTO substrate respectively, with and without light illumination.

**Figure 6.** ***Giant sharp and persistent converse magnetoelastic effects in BaTiO$_3$- based artificial multiferroic heterostructure via photo striction*****:** Isothermal Magnetization (at 100 K) versus time curves at the an applied a fixed value of magnetic field in the direction parallel to (100) direction of sample Ni/BaTiO$_3$ sample (a) at 100 Oe, (b) 400 Oe, The light pulse switching is shown in red step function curve; the height of step shows the On condition of the light pulse, otherwise denotes the OFF condition of light pulse, (c) The rate change of magnetization with the varying cycles at different

applied magnetic field, (d) X-ray magnetic circular dichroism with and without the illuminating the laser light of 532 nm.

**Figure 7. Direct evidence of collective domain wall Motion in the ferroelectric/ferromagnetic Ni/BaTiO$_3$ heterostructure:** Magneto-optic images of the magnetic domains of Ni/BaTiO$_3$ heterostructure with and without illumination of 405 nm laser light, In the present experiment the Dark contion is assumed as the illumination of sample during the MOKE measurement with the white light LED source. The dark initial and dark final are the MOKE image of Ni/BTO heterostructure before and after the visible light illumination.

**Figure.1**

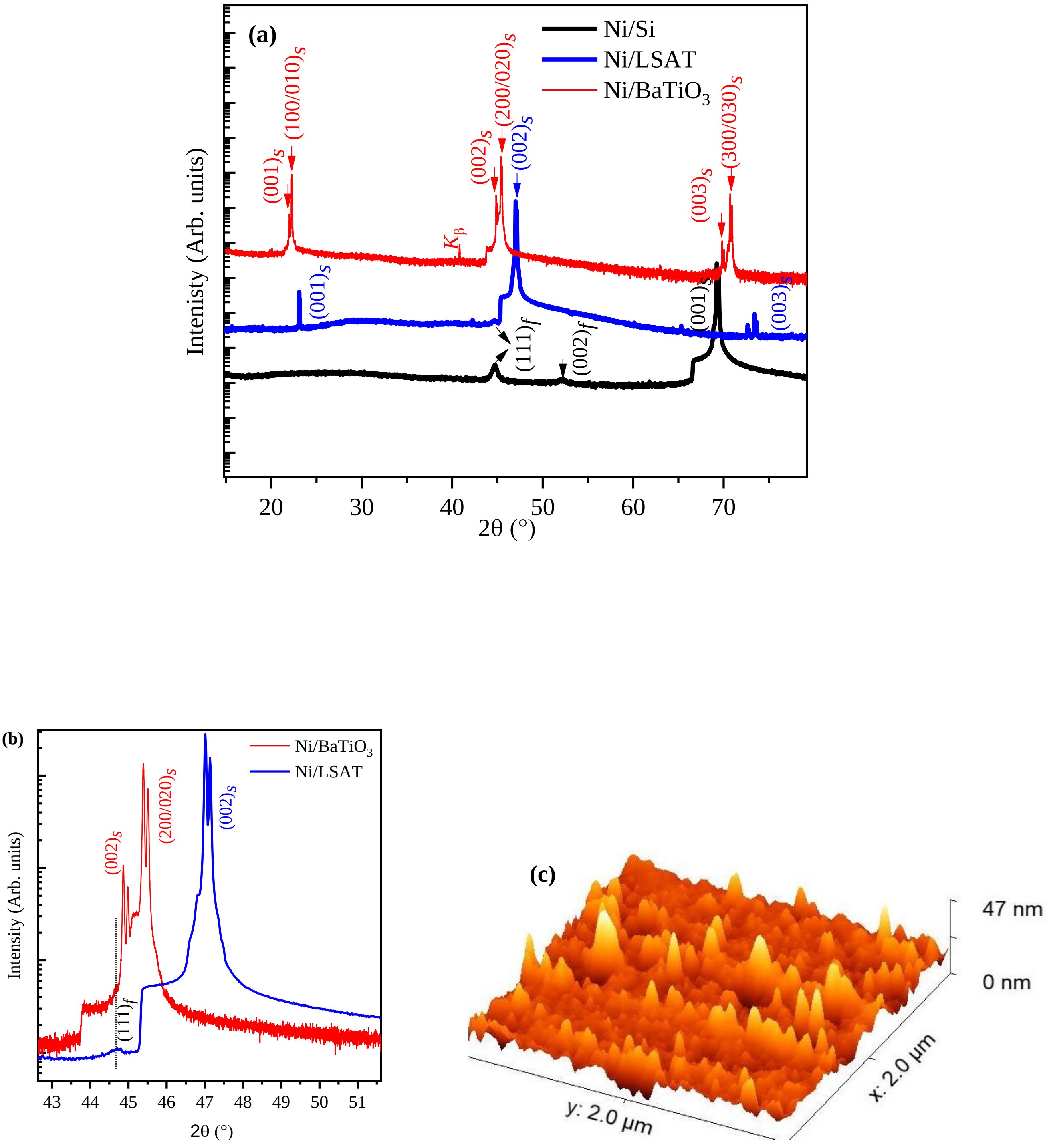

**Figure.2**

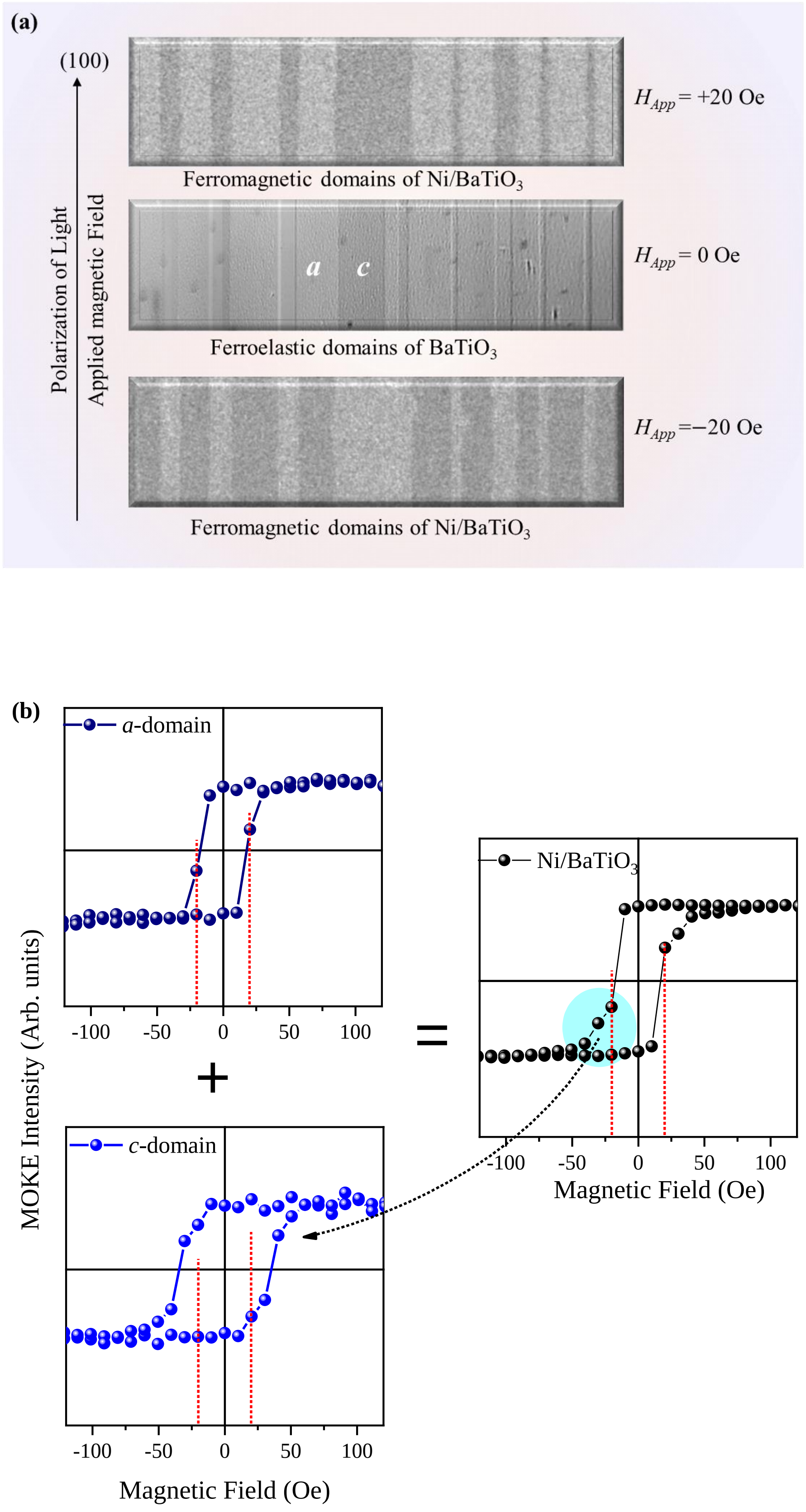

(a)
(100)
Polarization of Light
Applied magnetic Field
$H_{App}$ = +20 Oe
Ferromagnetic domains of Ni/$BaTiO_3$
a
c
$H_{App}$ = 0 Oe
Ferroelastic domains of $BaTiO_3$
$H_{App}$ = −20 Oe
Ferromagnetic domains of Ni/$BaTiO_3$
(b)
a-domain
c-domain
Ni/$BaTiO_3$
MOKE Intensity (Arb. units)
-100
-50
0
50
100
Magnetic Field (Oe)
+
=

**Figure.3**

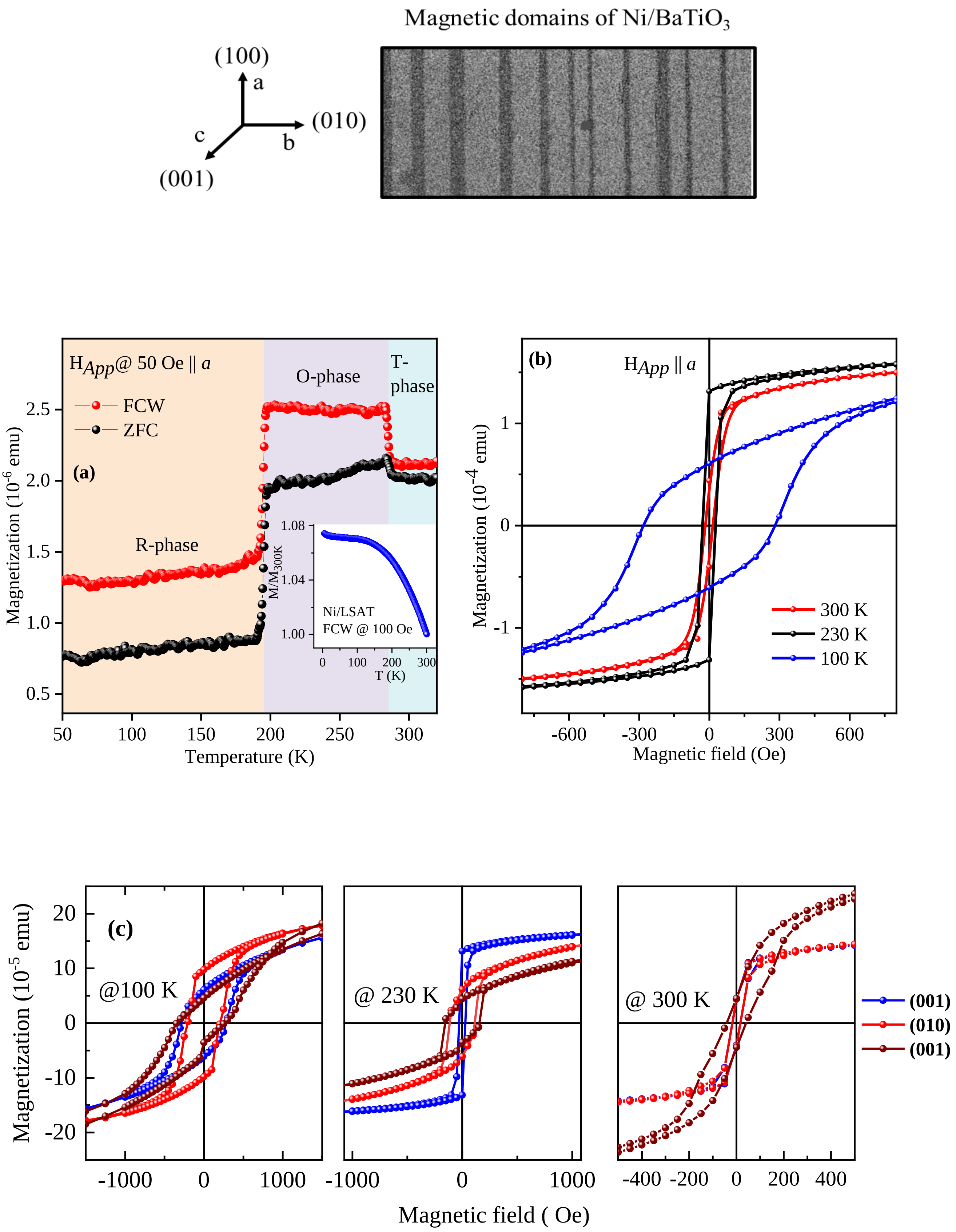

**Figure.4**

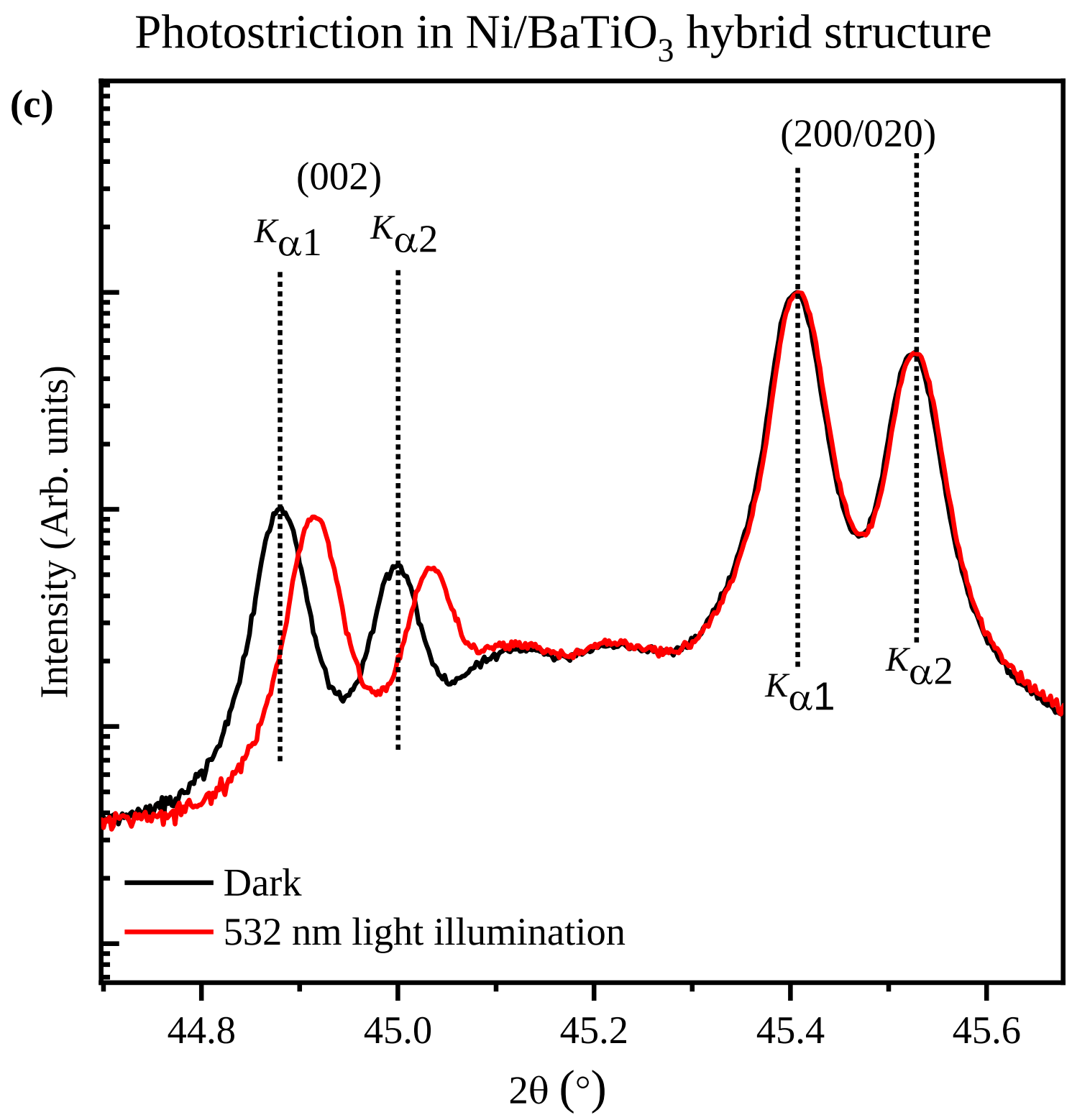

## Schematic I:

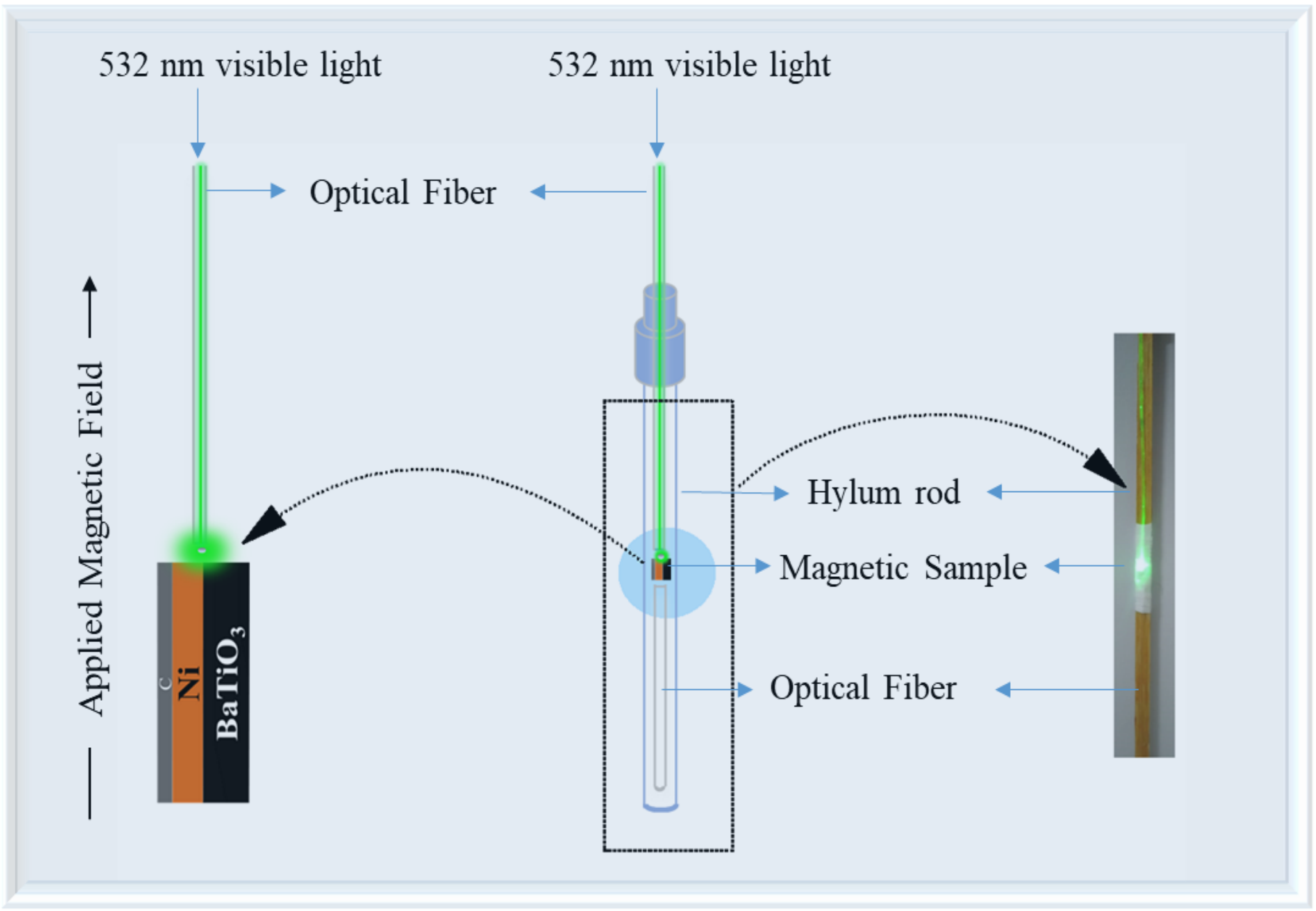

**Figure.5**

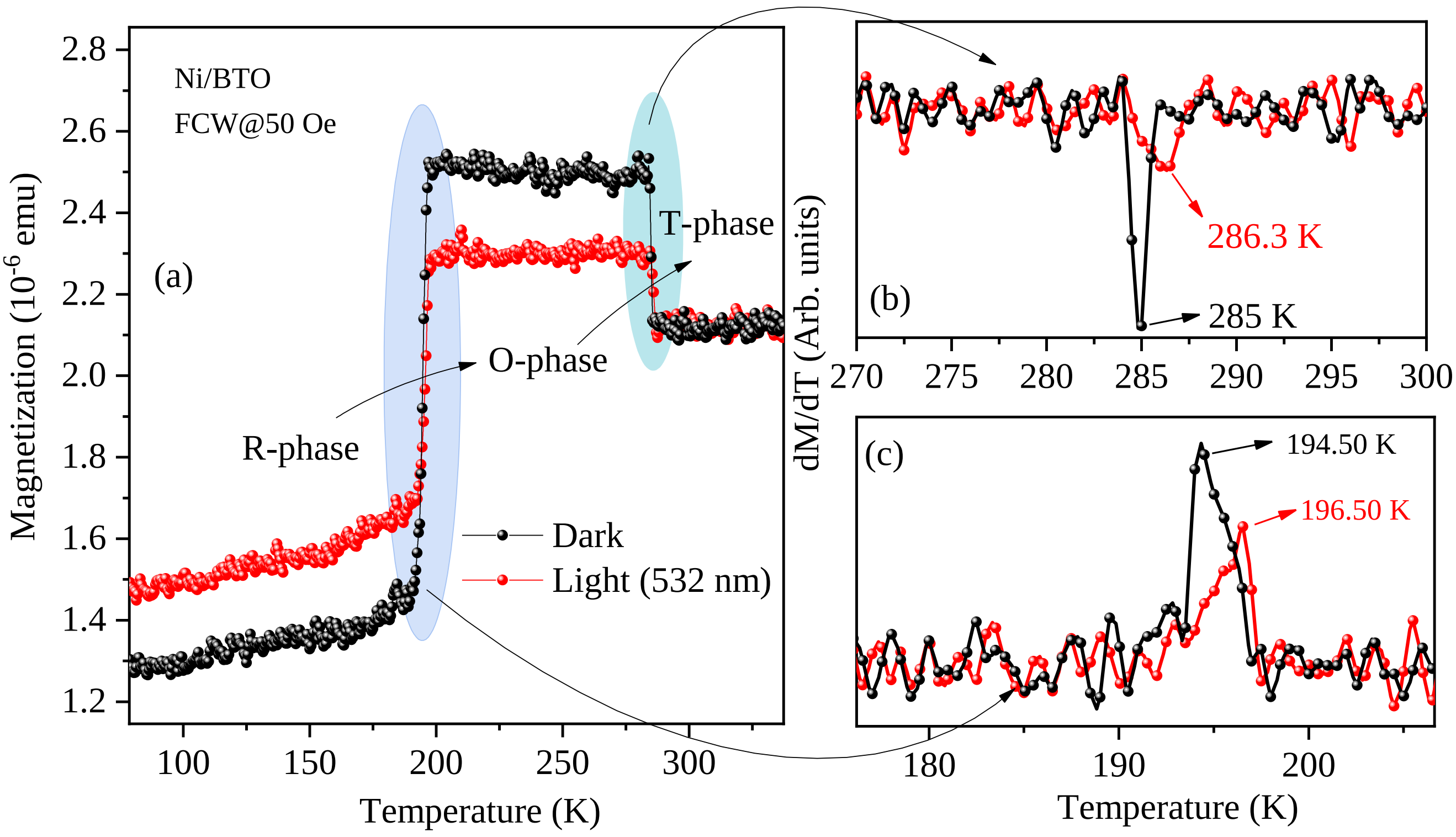

**Figure.6**

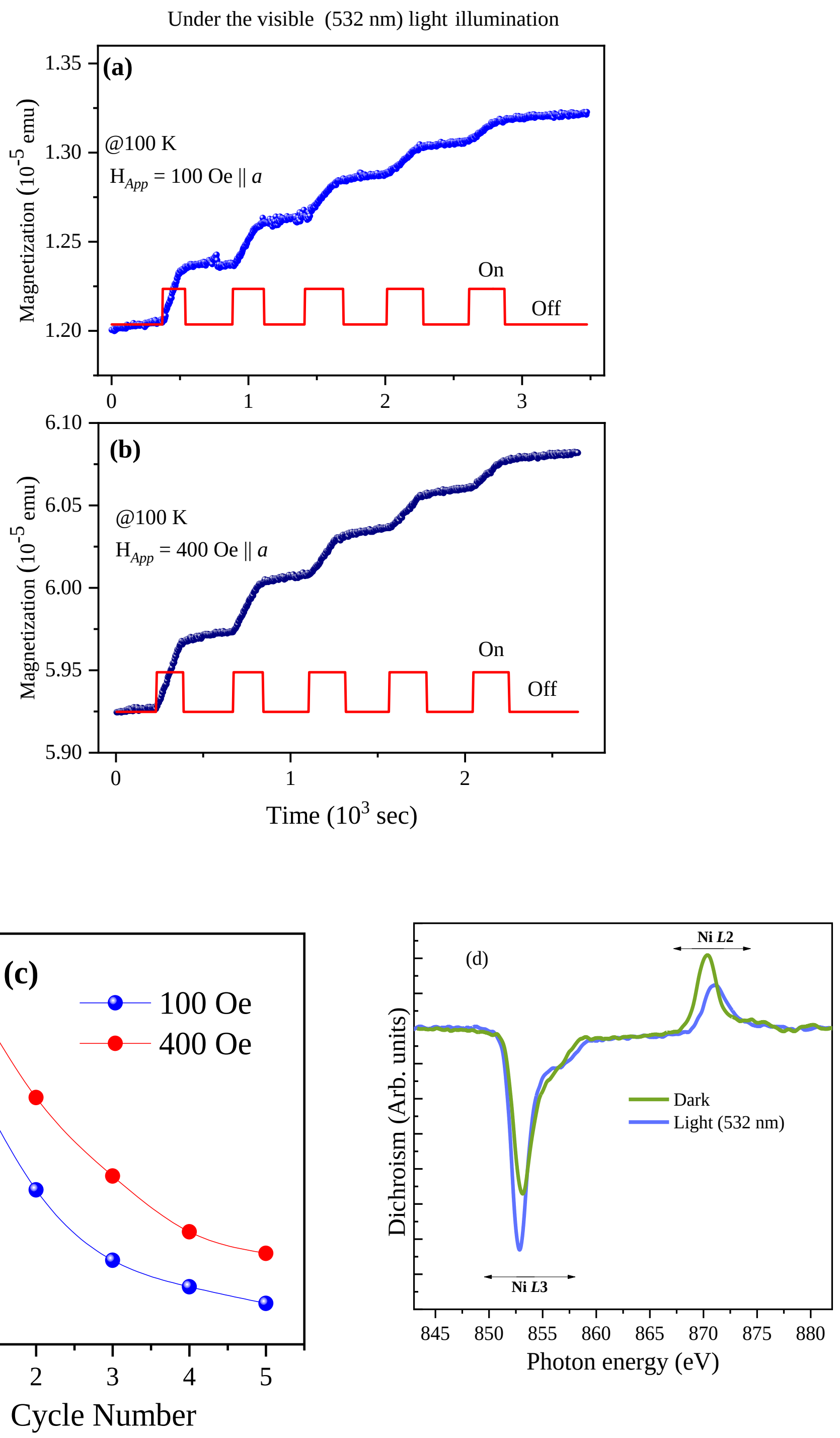

**Figure.7**

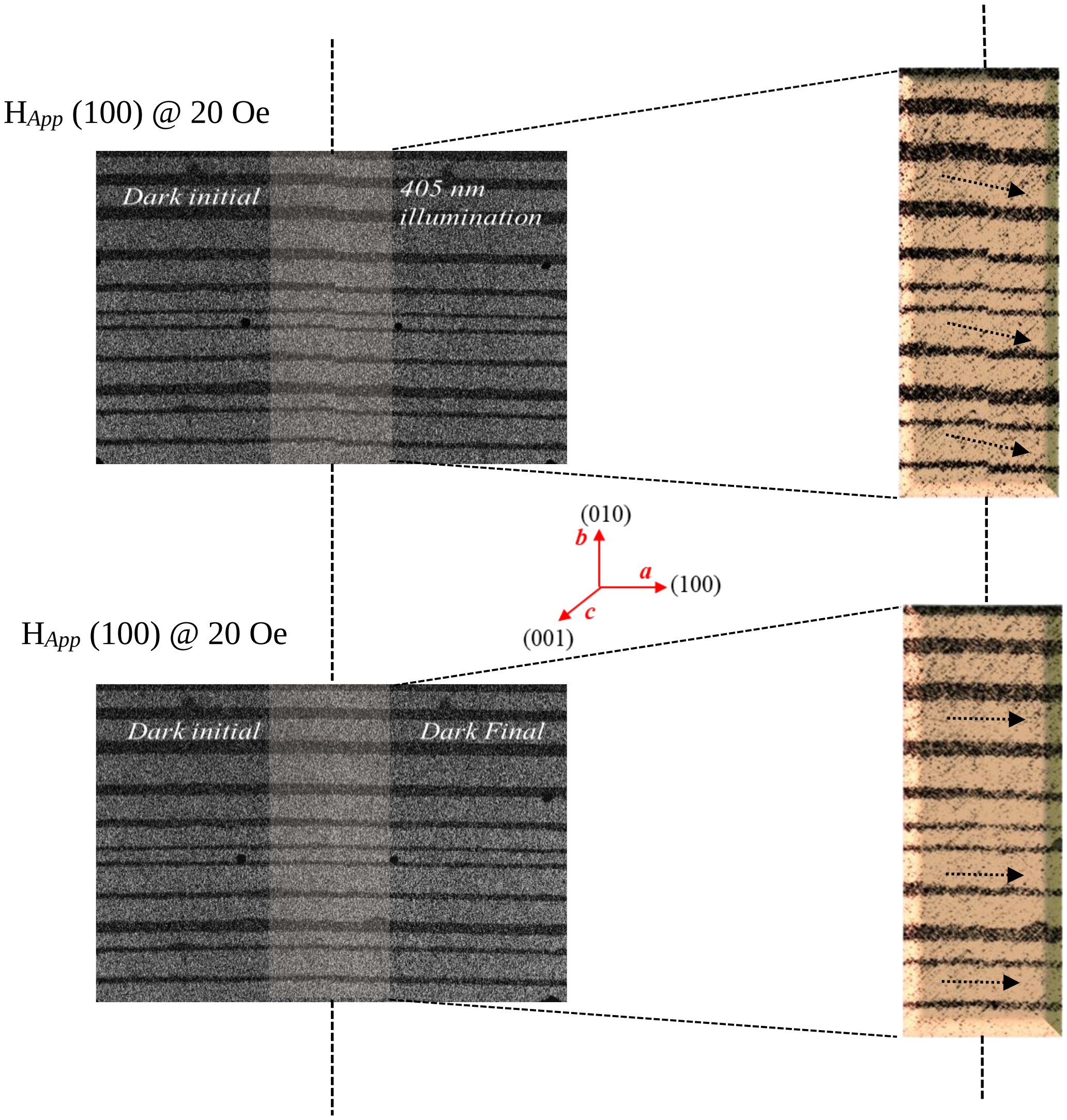